\documentclass[showpacs,aps,nofootinbib,floatfix,amsmath,amssymb]{revtex4}
\usepackage{graphicx}
\usepackage{relsize}
\begin{document}
\title{$U_e(1)$--covariant $R_\xi$--gauge for the two Higgs doublet model}

\author{C. G. Honorato$^{(a)}$ and J. J. Toscano$^{(a,b)}$}
\address{$^{(a)}$Facultad de Ciencias F\'{\i}sico Matem\'aticas,
Benem\'erita Universidad Aut\'onoma de Puebla, Apartado Postal
1152, Puebla, Puebla, M\'exico.\\
$^{(b)}$Instituto de F\'{\i}sica y Matem\' aticas, Universidad
Michoacana de San Nicol\' as de Hidalgo, Edificio C-3, Ciudad
Universitaria, C.P. 58040, Morelia, Michoac\' an, M\' exico.}

\begin{abstract}
An $U_e(1)$--covariant $R_\xi$ gauge for the two Higgs doublet
model based in the BRST symmetry is introduced. This gauge allows
one to remove an important number of nonphysical vertices
appearing in conventional linear gauges, which greatly simplifies
the loop calculations, since the the resultant theory satisfies
QED--like Ward identities. The presence of four ghost interactions
in this type of gauges and its connection with the BRST symmetry
is stressed. The Feynman rules for those new vertices that arise
in this gauge, as well as for those couplings already present in
the linear $R_\xi$ gauge but that are modified by this
gauge--fixing procedure, are presented.
\end{abstract}

\maketitle
\section{Introduction}
While gauge invariance plays a central role when defining the classical
action of a gauge system, once the latter is quantized one must invariably
invoke an appropriate gauge--fixing procedure to define a nondegenerate action,
which means that gauge invariance is to broken explicitly~\cite{LG}. The resultant
action is not gauge invariant, though it is invariant under BRST symmetry~\cite{BRST}.
As a consequence, the Green functions derived from this action cannot satisfy simple
 (QED--like) Ward identities, but they do satisfy more elaborated Slavnov--Taylor
identities that are dictated by BRST symmetry. Although in conventional quantization
schemes the quantum action of the theory is not gauge--invariant, it is still possible
to introduce gauge invariance with respect to a subgroup of such a theory. The simplest
 example of this quantization scheme is the nonlinear gauge introduced by K.
Fujikawa~\cite{NLGSM-Fuji} in the context of the electroweak
theory, which allows one to construct a quantum action that is
invariant under the electromagnetic gauge group. This
gauge--fixing procedure has been widely used in radiative
corrections, as it greatly simplifies the loop
calculations~\cite{ANLGSM}. More recently, this quantization
scheme has been generalized to include larger gauge groups than
the electroweak one. In Ref. \cite{NLG331}, one gauge--fixing
procedure covariant under the $SU_L(2)\times U_Y(1)$ group was
defined to quantize an electroweak extension based in the
$SU_L(3)\times U_X(1)$ gauge group~\cite{331}.

The main goal of this work is to introduce a general
renormalizable nonlinear gauge-fixing procedure for the two Higgs doublet model (THDM), which
is intended to remove the most nonphysical vertices from the
interaction Lagrangian, thereby facilitating the calculation of
radiative corrections considerably. The relevance of
nonlinear $R_\xi$ gauges \cite{NLGSM} in
radiative corrections has been emphasized by several authors not
only within the context of the standard model (SM) \cite{HHSM,NLGSM}, but in some of
its extensions such as the THDM \cite{NLGTHDM}, the so-called $331$
model \cite{NLG331}, and also in the model independent effective
Lagrangian approach \cite{NLGEFT}. In this paper we will discuss to
some extent the most general structure of a nonlinear $R_\xi$ gauge
for the THDM. We will argue that the Faddeev-Popov
method (FPM) \cite{FPM} fails when applied to this class of gauges
because the resultant theory is nonrenormalizable. Instead, the
appropriate framework is BRST symmetry \cite{BRST}, which is a
powerful formalism suited to quantize Yang-Mills theories with
broader supplementary conditions, and also  more general gauge
systems. To make clear this point, let us remind that the FPM leads
to an action which is bilinear in the ghost and antighost fields as
they arise essentially from the integral representation of a
determinant \cite{FPM}. However, this is not the most general
situation that can arise since an action including four-ghost
interactions at the tree level is still consistent with BRST
symmetry and the power counting criterion of renormalization theory.
It turns out that the FPM succeeds when used with conventional
linear gauges \cite{LG} because no four-ghost interaction can arise
from loop effects due to the existence of antighost translation
invariance, \footnote{Invariance under the transformation
$\bar{C}^a\to \bar{C}^a+c^a$, with $c^a$ an arbitrary constant
parameter} which in turn arises as a consequence of the fact that
the antighost fields appear just through their derivatives. However,
in the case of a nonlinear gauge, antighost translation invariance
is lost as the gauge-fixing functions depend on bilinear terms of
the gauge fields. These terms are responsible for the appearance of
ultraviolet-divergent four-ghosts interactions at the one-loop
level. This means that renormalizability becomes ruined when the
FPM is used in conjunction with nonlinear gauges. It is thus
convenient to desist from the use of the FPM and construct instead
the most general action consistent with BRST symmetry and
renormalization theory. Although these facts have been long known
from the study of Yang-Mills theories without spontaneous symmetry
breaking (SSB) \cite{YM}, they are less known in the context of the
SM and its extensions. Only more recently a complete Lagrangian for
the ghost sector of the SM was presented \cite{BT}. In this work we
pursue this issue and present a comprehensive study of nonlinear
$R_\xi$ gauges within the context of the THDM. Apart of eliminating much of the nonphysical vertices appearing in the THDM, our gauge--fixing procedure reduces in an appropriate limit to the corresponding nonlinear procedure for the SM given in Ref.\cite{BT}.

The paper is organized as follows. In Sec. \ref{model}, the main features of the THDM are discussed. Sec. \ref{gf} is devoted to present the most general nonlinear $R_\xi$ gauge for the THDM. Finally in Sec. \ref{c} the conclusions are presented.

\section{The model}
\label{model} We turn now to the main features of the Higgs sector
of the THDM. We will focus on the CP-conserving Higgs potential, which is necessary to contextualize the nonlinear
$R_\xi$ gauge. We will also emphasize the phenomenological
importance of the model and introduce the notation and conventions
used throughout the rest of the paper. The THDM incorporates two
scalar doublets of hypercharge $+1$:
$\Phi^\dag_1=(\phi^-_1,\phi_1^{0*})$ and
$\Phi^\dag_2=(\phi^-_2,\phi_2^{0*})$. The most general gauge
invariant potential can written as
\begin{eqnarray}
V(\Phi_1,\Phi_2)&=&\mu^2_1(\Phi_1^\dag
\Phi_1)+\mu^2_2(\Phi^\dag_2\Phi_2)-\left(\mu^2_{12}(\Phi^\dag_1\Phi_2)+{\rm
H.c.}\right)+
\lambda_1(\Phi^\dag_1\Phi_1)^2\nonumber \\
&&+\lambda_2(\Phi^\dag_2\Phi_2)^2+\lambda_3(\Phi_1^\dag
\Phi_1)(\Phi^\dag_2\, \Phi_2)+\lambda_4(\Phi^\dag_1\Phi_2)(\Phi^\dag_2\Phi_1)\nonumber \\
&&+
\frac{1}{2}\left(\lambda_5(\Phi^\dag_1\Phi_2)^2+\left(\lambda_6(\Phi_1^\dag
\Phi_1)+\lambda_7(\Phi^\dag_2\Phi_2)\right)(\Phi_1^\dag \Phi_2)+
{\rm H.c}.\right) \label{potential}
\end{eqnarray}

It is usual to impose the discrete symmetry $\Phi_1\to \Phi_1$ and
$\Phi_2\to -\Phi_2$ in order to avoid dangerous flavor changing
neutral current (FCNC) effects. This symmetry is strongly violated
by the $\lambda_6$ and $\lambda_7$ terms, but it is softly violated
by $\mu^2_{12}$. Nevertheless, all of these terms are essential to
obtain the decoupling limit of the model in which only one CP-even
scalar is light. As long as these violating terms exist, there are
two independent energy scales \cite{Gunion}, $v$ (the Fermi scale) and $\Lambda_{\rm
THDM}$, and the spectrum of Higgs boson masses is such that $m_{h}$
is of the order of $v$, whereas $m_{H}$, $m_{A}$ and $m_{H^\pm}$ are
all of the order of $\Lambda_{\rm THDM}$. In this case, all of the
heavy Higgs bosons decouple in the limit of $\Lambda_{\rm THDM}\gg
v$, according to the decoupling theorem. On the other hand, when the
scalar potential does respect the discrete symmetry, it is
impossible to have two independent energy scales \cite{Gunion}. As a
consequence, all of the physical scalar masses lie on the Fermi
scale $v$. Since $v$ is already fixed by the experiment, a very
heavy Higgs boson can only arise through a large dimensionless
coupling constant $\lambda_i$. In this scenario the decoupling
theorem is no longer valid, thereby opening the possibility for the
appearance of nondecoupling effects. In addition, since the scalar
potential contains some terms that violate the $SU(2)$ custodial
symmetry, nondecoupling effects can arise in one-loop induced Higgs
boson couplings \cite{Kanemura}.

The scalar potential (\ref{potential}) has to be diagonalized to
yield the mass-eigenstates fields. The charged components of the
doublets lead to a physical charged Higgs boson and the
pseudo-Goldstone boson associated with the $W$ gauge field:

\begin{eqnarray}
&&G^\pm_W=\phi^\pm_1c_\beta+\phi^\pm_2s_\beta,\\
&&H^\pm=-\phi^\pm_1s_\beta+\phi^\pm_2c_\beta,
\end{eqnarray}
with $\tan\beta=v_2/v_1$, being $v_1/\sqrt{2}\;\, (v_2/\sqrt{2})$
the vacuum expectation value (VEV) associated with
$\Phi_1\,(\Phi_2)$, and
$m^2_{H^\pm}=-(v^2_1+v^2_2)(\lambda_4+\lambda_5-2\mu^2_{12}/(v_1v_2))/2$.
Also, we have included the shorthand notation $s_\beta=\sin\beta$
and $c_\beta=\cos\beta$. On the other hand, the imaginary part of
the neutral components $\phi^0_{iI}$ defines the neutral CP-odd
scalar and the pseudo-Goldstone boson associated with the $Z$
gauge boson. The corresponding rotation is given by

\begin{eqnarray}
&&G_Z=\phi^0_{1I}c_\beta+\phi^0_{2I}s_\beta, \\
&&A=-\phi^0_{1I}s_\beta+\phi^0_{2I}c_\beta,
\end{eqnarray}
where $m^2_{A}=-(v^2_1+v^2_2)(\lambda_5-\mu^2_{12}/(v_1v_2))$.
Finally, the real part of the neutral components of the
$\phi^0_{iR}$ doublets defines the CP-even Higgs bosons $h$ and $H$.
The mass matrix is given by
\begin{equation}
M_{Re}=\left( \begin{array}{ccc} m_{11} & m_{12} \\
m_{12} & m_{22}\\
\end{array}\right),
\end{equation}
where

\begin{eqnarray}
&&m_{11}=2v^2_1\lambda_1+\frac{v_2}{v_1}\mu^2_{12},\\
&&m_{22}=2v^2_2\lambda_2+\frac{v_2}{v_1}\mu^2_{12}, \\
&&m_{12}=v_1v_2(\lambda_3+\lambda_4+\lambda_5)-\mu^2_{12}.
\end{eqnarray}

The physical CP-even states, $h$ and $H$, are given by

\begin{eqnarray}
&&H=\phi^0_{1R}c_\alpha+\phi^0_{2R}s_\alpha, \\
&&h=-\phi^0_{1R}s_\alpha+\phi^0_{2R}c_\alpha,
\end{eqnarray}

where

\begin{equation}
\tan 2\alpha=\frac{2m_{12}}{m_{11}-m_{22}},
\end{equation}
and
\begin{equation}
m^2_{H,h}=\frac{1}{2}\left(m_{11}+m_{22}\pm
\sqrt{(m_{11}-m_{22})^2+4m^2_{12}}\right).
\end{equation}

\section{The gauge-fixing procedure}
\label{gf} As already mentioned, the FPM fails to quantize
Yang-Mills theories with more general supplementary conditions than
the linear ones. To quantize Yang-Mills theories involving more
general gauge-fixing functions or either more general gauge systems,
it is thus necessary to invoke BRST symmetry. Although this symmetry
can be implemented in conventional field theory \cite{BRST,Ojima},
it arises more naturally within the context of the field-antifield
formalism \cite{FAF}. Therefore, in order to clarify our
presentation as much as possible, we will present a brief discussion
of this formalism. Although the following discussion is rather
general, we will focus on the properties of Yang-Mills systems. The
starting point  is the introduction of an antifield for each field
present in the theory. It is assumed that the dynamical degrees of
freedom of the gauge system are characterized by the matter, gauge,
ghost ($C^a$), antighost ($\bar{C}^a$), and auxiliary ($B^a$)
fields. The original action, which will be denoted by $S_0$, is a
functional of the matter and the gauge fields only, but this
configuration is extended to include the ghost fields because they
are necessary to quantize the theory. A ghost field for each gauge
parameter is introduced. The ghost fields have opposite statistics
to that of the gauge parameters. To gauge fix and quantize the
theory, it is necessary to introduce the so-called trivial pairs,
namely the antighost and auxiliary fields. We let $\Phi^A$ run over
all these fields. For each $\Phi^A$, an antifield $\Phi^*_A$ is
introduced, with opposite statistics to $\Phi^A$ and a ghost number
equal to $-gh(\Phi^A)-1$, where $gh(\Phi^A)$ is the ghost number of
$\Phi^A$. It is $0$ for matter, gauge, and auxiliary fields, $+1$
for ghosts, and $-1$ for antighosts. In this extended configuration
space a symplectic structure is introduced through left and right
differentiation, defined for two functionals $F$ and $G$ as:
\begin{equation}
(F,G)=\frac{\delta_RF}{\delta \Phi^A}\frac{\delta_LG}{\delta
\Phi^*_A}-\frac{\delta_RF}{\delta \Phi^*_A}\frac{\delta_LG}{\delta
\Phi^A}.
\end{equation}
In particular, the fundamental antibrackets are given by
\begin{eqnarray}
&&(\Phi^A,\Phi^*_B)=\delta^A_B, \\
&&(\Phi^A,\Phi^B)=(\Phi^*_A,\Phi^*_B)=0.
\end{eqnarray}
The extended action is a bosonic functional of the fields and the
antifields, $S[\Phi,\Phi^*]$, with ghost number zero, which
satisfies the master equation defined by
\begin{equation}
(S,S)=2\frac{\delta_RS}{\delta \Phi^A}\frac{\delta_LS}{\delta
\Phi^*_A}=0.
\end{equation}
The antibrackets serve to define the extended BRST transformations:
\begin{eqnarray}
&&\delta_B\Phi^A=(S,\Phi^A)=-\frac{\delta_R S}{\delta \Phi^*_A},
\\
&&\delta_B\Phi^*_A=(S,\Phi^*_A)=\frac{\delta_RS}{\delta \Phi^A}.
\end{eqnarray}
Thus, the BRST transformations are generated by the extended action,
which is invariant under these transformations due to the master
equation as its variation is  $\delta_BS=(S,S)$. On the other hand,
not all the solutions of the master equation are of interest, but
only those called proper solutions \cite{FAF}. A proper solution
must make contact with the initial theory, which means to impose the
following boundary condition on $S$:
\begin{equation}
S[\Phi,\Phi^*]|_{\Phi^*=0}=S_0[\phi],
\end{equation}
where $\phi$ runs only over the original fields, i.e., matter and
gauge fields. The proper solution $S$ can be expanded in a power
series in antifields:
\begin{equation}
S[\Phi, \Phi^*]=S_0[\phi]+(\delta_B\Phi^A)\Phi^*_A+\cdots,
\end{equation}
in which all the gauge-structure tensors characterizing the gauge
system appear. In this sense, the proper solution $S$ is the
generating functional of the gauge-structure tensors. $S$ also
generates the gauge algebra through the master equation. So,
classically a gauge system is completely determined when the proper
solution $S$ is established and the master equation is calculated,
which yields the relations that must be satisfied by the
gauge-structure tensors. In the simplest gauge systems, such as
Yang-Mills theories, a solution of the master equation is given by
\begin{equation}
S[\Phi,\Phi^*]=S_0[\phi]+(\delta_B\Phi^A)\Phi^*_A.
\end{equation}
This action is bosonic and has ghost number zero as required. It is
easy to show that this action is a solution of the master equation
and accurately reproduces the well-known gauge algebra of Yang-Mills
theories.

We turn now to the quantum analysis of the gauge system. To quantize
the theory, one starts by fixing the gauge. Since the extended
action is degenerate, it cannot be quantized directly. Furthermore,
the antifields do not represent true degrees of freedom, so they
must be removed before quantizing the theory. However they cannot be
just set to zero since $S_0$ is degenerate. One can remove the
antifields instead through a nontrivial procedure and at the same
time lift the degeneration of the theory. The antifields can be
eliminated by introducing a fermionic functional of the fields,
$\Psi [\Phi]$, with ghost number $-1$, such that
\begin{equation}
\Phi^*_A=\frac{\delta \Psi[\Phi]}{\delta\Phi^A}.
\end{equation}
Note that it is not necessary to distinguish between left- and
right-differentiation. In defining a gauge-fixing procedure, the
presence of the trivial pairs, $\bar{C}^a$ and $B^a$, is necessary
since the only fields with ghost number $-1$ are precisely the
antighosts. By noting that
$(\delta_B\Phi^A)\Phi^*_A=(\delta_B\Phi^A)(\delta \Psi[\Phi]/\delta
\Phi^A=\delta\Psi[\Phi]$, the proper solution takes the form
\begin{equation}
S[\Phi,\delta \Psi/\delta\Phi]=S_0[\phi]+\delta\Psi[\Phi].
\end{equation}
This is the gauge-fixed BRST action, which is invariant under the
well-known BRST transformations \cite{BRST}. From now on, we will
denote these transformations by the symbol $\delta$. It should be
noted that, in contrast with the extended action $S$, $S_0+\delta
\Psi[\Phi]$ is not degenerate. In general, the nilpotency of
$\delta$ is only guaranteed on-shell, i.e., only after using the
equations of motion, but in the case of Yang-Mills theories,
$\delta^2=0$ even off-shell.

We now proceed to define the most general fermionic functional
$\Psi[\Phi]$ for the electroweak group $SU_L(2)\times U_Y(1)$,
consistent with renormalization theory.  The most general $\Psi$
functional with ghost number $-1$ can be written as follows:
\begin{equation}
\Psi=\int
d^4x\Big[\bar{C}^a\Big(f^a+\frac{\xi}{2}B^a+g\epsilon^{abc}\bar{C}^bC^c\Big)
+\bar{C}\Big(f+\frac{\xi}{2}B\Big)\Big],
\end{equation}
where $f^a$ and $f$ are the gauge-fixing functions associated with
the groups $SU_L(2)$ and $U_Y(1)$, respectively. They are restricted
by renormalizability to be at most quadratic functions of the gauge
and the scalar fields. The bosonic constant $\xi$ is the so-called
gauge parameter. In general there is one such parameter for each
group, but we have used the same by simplicity. $B^a$ and $B$ are
the auxiliary fields associated with the groups $SU_L(2)$ and
$U_Y(1)$, respectively. Note that the term
$g\epsilon^{abc}\bar{C}^a\bar{C}^bC^c$ cannot exist in the
Faddeev-Popov approach, though its presence is necessary to get
renormalizability when the gauge-fixing functions are nonlinear.
Using the usual BRST transformations, we obtain for the $\Psi$
variation
\begin{eqnarray}
\delta\Psi&=&\int d^4x
\Big\{\frac{\xi}{2}B^aB^a+g\Big(f^a+2\epsilon^{abc}\bar{C}^bC^c\Big)B^a+\frac{\xi}{2}BB+fB
\nonumber \\
&&-\bar{C}^a(\delta_{B_\Psi}f^a)
-\bar{C}(\delta_{B_\Psi}f)-g^2\bar{C}^a\bar{C}^bC^aC^b\Big\}.
\end{eqnarray}
On the other hand, since the auxiliary fields $B^a$ and $B$ appear
quadratically, they can be integrated out. Since the coefficients of
the quadratic terms do not depend on the fields, their integration
is equivalent to use the corresponding equations of motion in the
gauge-fixed BRST action. Once this is done, we obtain an effective
action defined by the following effective Lagrangian
\begin{equation}
{\cal L}_{eff}={\cal L}_{THDM}+{\cal L}_B+{\cal L}_F,
\end{equation}
where ${\cal L}_{THDM}$ is the gauge invariant Lagrangian of the
THDM, whereas ${\cal L}_B$ and ${\cal L}_F$ arise from the action
$\delta\Psi$. ${\cal L}_B$ is the gauge-fixing term, which can be
written as
\begin{equation}
{\cal L}_B=-\frac{1}{2\xi}f^af^a-\frac{1}{2\xi}(f)^2.
\end{equation}
On the other hand, ${\cal L}_F$ represents the ghost sector and it
is given by
\begin{eqnarray}
{\cal L}_F&=&-\bar{C}^a(\delta f^a)-\bar{C}(\delta
f)\nonumber \\
&&-\frac{2g}{\xi}\epsilon^{abc}
f^a\bar{C}^bC^c+g^2\Big(\frac{2}{\xi}-1\Big)\bar{C}^a\bar{C}^bC^aC^b.
\end{eqnarray}
While the first two terms correspond to those arising from the FPM,
the last two ones do not appear when this framework is used. It is
important to note that the $2g/\xi\,\epsilon^{abc} f^a\bar{C}^bC^c$
term can contribute to some class of Green functions yet at the
one-loop level, a fact to be considered with dealing with radiative
corrections.

\subsection{The nonlinear gauge-fixing functions}

We are now ready to discuss the gauge-fixing functions $f^a$ and $f$
in the context of the THDM. Our main aim is to remove the most
unphysical vertices that are generated by the Higgs kinetic-energy
term. We will take advantage of the fact that every coupling
involving at least one pseudo-Goldstone boson can be modified or
removed from the theory leaving unaltered the $S$ matrix. Also,
gauge freedom allows us to modify the Yang-Mills sector in a
nontrivial way. Bearing this in mind, the first step in our strategy
consist in defining a gauge-fixing procedure for the charged $W$
gauge field covariant under the $U_e(1)$ group. This means that the
covariant derivative associated with this group must be used instead
of the ordinary one. An adequate extension of this derivative allows
us to eliminate not only the unphysical vertex $WG_W\gamma$ but also
the $WG_WZ$ one, and simultaneously guarantees a highly symmetric
behavior of the theory. The second stage in our program consists in
introducing gauge-fixing functions nonlinear in the scalar sector,
meant to remove various unphysical quartic vertices, which involve
interactions between gauge bosons, pseudo-Goldstone bosons, and
physical scalar bosons. We thus propose the following gauge-fixing
functions for the THDM:
\begin{eqnarray}
&&f^a=f^a_V+f^a_S, \\
&&f=f_V+f_S,
\end{eqnarray}
where
\begin{eqnarray}
f^a_V&=&\Big(\delta^{ab}\partial_\mu-g'\epsilon^{3ab}B_\mu\Big)W^{b\mu},
\\
f^a_S&=&\frac{ig\xi}{2}\Big\{\sum_{i=1}^2\Big[\Phi^\dag_i(\sigma^a-i\epsilon^{3ab}\sigma^b)\Phi_{0i}-
\Phi^\dag_{0i}(\sigma^a+i\epsilon^{3ab}\sigma^b)\Phi_i\Big]\nonumber \\
&&+i\epsilon^{3ab}(c_\beta\Phi^\dag_1+s_\beta\Phi^\dag_2)\sigma^b(c_\beta
\Phi_1+s_\beta \Phi_2)\Big\},
\end{eqnarray}
and
\begin{eqnarray}
&&f_V=\partial_\mu B^\mu, \\
&&f_S=\frac{ig'\xi}{2}\sum_{i=1}^2\Big(\Phi^\dag_i\Phi_{0i}-\Phi^\dag_{0i}\Phi_i\Big).
\end{eqnarray}
In the above expressions, $\Phi^\dag_{0i}=(0,v_i/\sqrt{2})$,
$\sigma^a$ are the Pauli matrices, and $W^a_\mu$ and $B_\mu$ are the
gauge fields associated with the electroweak group. Our gauge-fixing
functions contain the conventional linear functions as a particular
case. In fact, they are obtained when $\epsilon^{3ab}$ is set to
zero. All these functions are Hermitian as required. Also, we have
introduced the linear combination of Higgs doublets $c_\beta
\Phi_1+s_\beta \Phi_2$, which is necessary in order to avoid the
presence of terms involving only physical fields. In other words,
any term appearing in $f^a_S$ and $f_S$ involves the presence of at
least one unphysical scalar. It is worth mentioning that this
gauge-fixing procedure contains as a particular case an analogous
gauge scheme for the minimal SM \cite{BT}, which becomes evident
when the $\Phi_1$ doublet is associated with the SM one and $\beta$
is set to zero

To fully appreciate the structure of the gauge-fixing functions, it
is convenient to express them in terms of mass eigenstates fields.
To this end, we use the following definitions:
\begin{eqnarray}
&&f^\pm=\frac{1}{\sqrt{2}}(f^1\mp i f^2),\\
&&f^Z=c_Wf^3-s_Wf,\\
&&f^A=s_Wf^3+c_Wf,
\end{eqnarray}
where $s_W\,(c_W)$ is the sine (cosine) of the weak angle. We then
obtain for the vector sector
\begin{eqnarray}
&&f^+_V=\bar{D}_\mu W^{+\mu}, \\
&&f^Z_V=\partial_\mu Z^\mu, \\
&&f^A_V=\partial_\mu A^\mu,
\end{eqnarray}
and for the scalar sector
\begin{eqnarray}
&&f^+_S=-\frac{ig\xi}{2}\Big(\varphi^0-iG_Z\Big)G^+_W,
\\
&&f^Z_S=-\xi m_ZG_Z, \\
&&f^A_S=0,
\end{eqnarray}
where $\varphi^0=v+c_{\beta-\alpha}H+s_{\beta-\alpha}h$ and
$\bar{D}_\mu=\partial_\mu-ig'B_\mu$, being $g'$ the coupling
constant associated with the $U_Y(1)$ group. We can see that both
$f^+_V$ and $f^+_S$ are nonlinear and transform covariantly under
the $U_e(1)$ group, as $\bar{D}_\mu$ contains the covariant
derivative associated with this group.

The gauge-fixing Lagrangian, ${\cal L}_B$, can then be written as
\begin{equation}
{\cal L}_B={\cal L}_{BV}+{\cal L}_{BS}+{\cal L}_{BSV},
\end{equation}
where
\begin{eqnarray}
{\cal
L}_{BV}&=&-\frac{1}{\xi}f^-_Vf^+_V-\frac{1}{2\xi}(f^Z_V)^2-\frac{1}{2\xi}(f^A_V)^2
\nonumber \\
&=&-\frac{1}{\xi}(\bar{D}_\mu W^{+\mu})^\dag (\bar{D}_\nu
W^{+\nu})-\frac{1}{2\xi}(\partial_\mu
Z^\mu)^2-\frac{1}{2\xi}(\partial_\mu A^\mu)^2,\\
{\cal L}_{BS}&=&-\frac{1}{\xi}f^-_Sf^+_S-\frac{1}{2\xi}(f^Z_S)^2
\nonumber \\
&=&-\frac{g^2\xi}{4}\Big(\varphi^{2}+G^2_Z\Big)G^-_WG^+_W
-\frac{1}{2}\xi m^2_ZG^2_Z,\\
{\cal
L}_{BSV}&=&-\frac{1}{\xi}\Big(f^-_Vf^+_S+f^-_Sf^+_V\Big)-\frac{1}{2\xi}f^Z_Vf^Z_S
\nonumber \\
&=&\frac{ig}{2}\Big[(\bar{D}_\mu W^{+\mu})^\dag
(\varphi-iG_Z)G^+_W-(\bar{D}_\mu
W^{+\mu})(\varphi+iG_Z)G^-_W\Big]+m_ZG_Z\partial_\mu Z^\mu.
\end{eqnarray}

We now proceed to investigate the consequences of these gauge-fixing
terms on the gauge invariant Lagrangian of the THDM.  Feynman rules for the THDM in conventional linear gauges have been discussed by several authors~\cite{Gunion,1Couture,2Mendez,3kingman,4arhrib,6malinsky}. Here, we will present only those couplings which are affected by our nonlinear gauge--fixing procedure. First of all,
the term ${\cal L}_{BV}$ defines the propagators of the gauge fields
and also modifies nontrivially the Lorentz structure of the
trilinear and quartic vertices arising from the Yang-Mills sector.
To clarify this point, it is convenient to write down the sum of
these two terms:
\begin{eqnarray}
\label{ym}
{\cal L}_V&=& {\cal L}_{YM}+{\cal L}_{BV} \nonumber \\
&=&-\frac{1}{2}(\hat{D}_\mu W^+_\nu-\hat{D}_\nu W^+_\mu)^\dag
(\hat{D}^\mu
W^{+\nu}-\hat{D}^\nu W^{+\mu}) \nonumber \\
&&-\frac{1}{4}Z_{\mu \nu}Z^{\mu \nu}-\frac{1}{4}F_{\mu \nu}F^{\mu
\nu} \nonumber \\
&&-ig\Big[s_WF_{\mu \nu}+c_WZ_{\mu \nu}+\frac{ig}{2}(W^-_\mu
W^+_\nu-W^-_\nu
W^+_\mu)\Big]W^{-\mu}W^{+\nu} \nonumber \\
&&-\frac{1}{\xi}(\bar{D}_\mu W^{+\mu})^\dag (\bar{D}_\nu
W^{+\nu})-\frac{1}{2\xi}(\partial_\mu
Z^\mu)^2-\frac{1}{2\xi}(\partial_\mu A^\mu)^2,
\end{eqnarray}
where $\hat{D}_\mu=\partial_\mu-igW^3_\mu$, being $g$ the constant
coupling associated with the $SU_L(2)$ group. It is thus evident
that, with the exception of the $WWWW$ vertex, all trilinear and
quartic vertices are modified by the gauge fixing procedure. Since
the term that introduces the modifications on these vertices is
invariant under the $U_e(1)$ group, the trilinear electromagnetic
vertices satisfy QED-like Ward identities. This fact is relevant for
radiative corrections as such a symmetry greatly simplifies this
class of calculations. On the other hand, as it can be appreciated from the above Lagrangian, the trilinear $WWV$ ($V=\gamma,Z$) vertex is modified in a nontrivial way by this gauge--fixing procedure. With the exception of the $WWWW$ vertex, the quartic interactions $WWVV$ are modified too. They now are gauge depending, \textit{i.e.}, they depend on the $\xi$ gauge parameter. The vertex functions associated with the $V^\mu(k_1)W^{+\lambda}(k_2)W^{-\rho}(k_3)$ and $V^\mu_1V^\nu_2W^{+\lambda}W^{-\rho}$ couplings are given by $ig_V\Gamma_{\lambda \rho \mu}(k_1,k_2,k_3)$ and $ig_{V_1}g_{V_2}\Gamma_{\lambda \rho \mu \nu}$, respectively, where
\begin{eqnarray}
\Gamma_{\lambda \rho \mu}(k_1,k_2,k_3)&=&(k_2-k_3)_\mu
g_{\lambda \rho}+(k_3-k_1+\frac{\delta_V}{\xi}k_2)_\lambda g_{\rho
\mu}\nonumber \\
&&+(k_1-k_2-\frac{\delta_V}{\xi}k_3)_\rho g_{\lambda \mu},
\end{eqnarray}
\begin{equation}
\Gamma_{\lambda \rho \mu \nu}=-2g_{\mu \nu}g_{\lambda \rho}+\left(1-\frac{\delta_{V_1V_2}}{\xi}\right)\left(g_{\lambda \mu}g_{\rho \nu}+g_{\rho \mu}g_{\lambda \nu}\right),
\end{equation}
where all momenta are token incoming. In addition, we have introduced the following definitions:
\begin{equation}
\delta_{V_1V_2}=\left\{ \begin{array}{lll} +1,
& V_1V_2=\gamma \gamma\\
-\frac{s^2_W}{c^2_W}, & V_1V_2=\gamma Z,\\
+\frac{s^4_W}{c^4_W}, & V_1V_2=ZZ,
\end{array} \right.
\end{equation}

\begin{equation}
g_V=\left\{ \begin{array}{ll} gs_W,
& V=\gamma\\
gc_W, & V=Z.
\end{array} \right.
\end{equation}

As for the ${\cal L}_{BS}$ term, it defines the masses of the $G_W$
and $G_Z$ fields and modifies some unphysical couplings arising from
the Higgs potential. In this gauge, the couplings between scalar
fields arise solely  from the sum of the following terms
\begin{eqnarray}
\label{hp}
{\cal L}_S&=&-V(\Phi_1,\Phi_2)+{\cal L}_{BS} \nonumber \\
&=&-V(\Phi_1,\Phi_2)-\frac{1}{2}\xi
m^2_ZG^2_Z-\frac{g^2\xi}{4}(\varphi^{02}+G^2_Z)G^-_WG^+_W.
\end{eqnarray}
From this expression, it is clear that the gauge-fixing procedure
modifies the strength of the couplings $HG^-_WG^+_W$, $hG^-_WG^+_W$,
$H^{2}G^-_WG^+_W$, $h^{2}G^-_WG^+_W$, $HhG^-_WG^+_W$, and
$G^2_ZG^-_WG^+_W$. The modified Feynman rules are shown in Table \ref{TABLE1}. The physical couplings remain unchanged, as
required.

\begin{table}
 \caption{\label{TABLE1} Nonphysical couplings of the Higgs
 potential that are modified by the nonlinear gauge--fixing
 procedure.}
 \begin{tabular}{|l|l|}
 \hline
 Coupling & Vertex Function  \\
 \hline $HHG_W^+G_W^-$&$\frac{-ig^2}{4m_W^2}\left(2m_{H^\pm}^2
 s_{\alpha+\beta}^2-\frac{m_h^2s_{2\alpha}s_{\alpha-\beta}^2}
 {s_{2\beta}}+\frac{m_H^2(s_{\beta-3\alpha}+3s_{\alpha+\beta})
 c_{\alpha-\beta}}{2s_{2\beta}} +2m_W^2\xi
 c_{\alpha-\beta}^2\right)$\\
 \hline
 $hhG_W^+G_W^-$&$\frac{-ig^2}{4m_W^2}\left(2m_{H^\pm}c_{\alpha-\beta}^
 2+\frac{m_H^2s_{2\alpha}c_{\alpha-\beta}^2}{s_{2\beta}}
 -\frac{m_h^2(c_{\beta-3\alpha}+3c_{\alpha+\beta})
 s_{\alpha-\beta}}{2s_{2\beta}}+2m_W^2\xi
 s_{\alpha-\beta}^2\right)$\\
 \hline
 $G_ZG_ZG_W^+G_W^-$&$\frac{-ig^2}{4m_W^2}\left(m_H^2c_{\alpha-\beta}^2
 +m_h^2s_{\alpha-\beta}^2+2m_W^2\xi\right)$\\
 \hline
 $HhG_W^+G_W^-$&$\frac{-ig^2}{4m_W^2}\left(m_{H^\pm}^2s_{2(\alpha-\beta)}
 -\frac{(m_h^2-m_H^2)s_{2\alpha}s_{2(\alpha-\beta)}}{2s_{2\beta}}-m_W^2\xi
 s_{2(\alpha-\beta)}\right)$\\
 \hline $HG_W^+G_W^-$&$\frac{-ig}{2m_W}\left(m_H^2+2m_W\xi
 \right)c_{\beta-\alpha}$\\
 \hline $hG_W^+G_W^-$&$\frac{-ig}{2m_W}\left(m_h^2+2m_W\xi
 \right)s_{\beta-\alpha}$\\
 \hline
 \end{tabular}
\end{table}

We now turn to discuss the dynamical implications of the term ${\cal
L}_{BSV}$. This term affects considerably the Higgs kinetic-energy
sector of the theory since it removes several unphysical vertices.
When these two terms are combined, we obtain
\begin{eqnarray}
\label{hk} &&\sum^2_{i=1}(D_\mu \Phi_i)^\dag (D^\mu \Phi_i)+{\cal
L}_{BSV}=\frac{1}{2}\sum_{\phi=h,H,A,G_Z}(\partial_\mu
\phi)(\partial^\mu \phi)+\sum_{\phi^+=H^+,G^+_W}(\partial_\mu
\phi^-)(\partial^\mu \phi^+)\nonumber \\
&&+\frac{ig}{2}\{W^-_\mu [-2G^+_W\partial^\mu
S_1+2iG^+_W\partial^\mu G_Z+S_2\partial^\mu H^+-H^+\partial^\mu
S_2 \nonumber \\
&&+i(H^+\partial^\mu A-A\partial^\mu
H^+)-ie(A^\mu-\frac{s_W}{c_W}Z^\mu)(S_2-iA)H^+]-h.c.\}\nonumber \\
&&+\frac{g}{2c_W}Z_\mu [c_{\beta-\alpha}(S_2\partial^\mu
P_1-P_1\partial^\mu S_2+S_1\partial^\mu P_2-P_2\partial^\mu
S_1)\nonumber \\
&&+s_{\beta-\alpha}(S_1\partial^\mu P_1-P_1\partial^\mu
S_1+P_2\partial^\mu S_2-S_2\partial^\mu P_2)]\nonumber \\
&&+ie(A_\mu+\frac{c_{2W}}{s_{2W}}Z_\mu)(G^-_W\partial^\mu
G^+_W-G^+_W\partial^\mu G^-_W+H^-\partial^\mu H^+-H^+\partial^\mu
H^-)\nonumber \\
&&+e^2(A_\mu A^\mu+2\frac{c_{2W}}{s_{2W}}A_\mu
Z^\mu+\frac{c^2_{2W}}{s^2_{2W}}Z_\mu Z^\mu+\frac{1}{2s^2_W}W^-_\mu
W^{+\mu})(G^-_WG^+_W+H^-H^+)\nonumber \\
&&+\frac{g^2}{4}(W^-_\mu W^{+\mu}+\frac{1}{2c^2_W}Z_\mu
Z^\mu)(v^2+2vS_1+h^2+H^2+A^2+G^2_Z),
\end{eqnarray}
where
\begin{eqnarray}
&&S_1=c_{\beta-\alpha}H+s_{\beta-\alpha}h, \\
&&S_2=-s_{\beta-\alpha}H+c_{\beta-\alpha}h,
\end{eqnarray}
\begin{eqnarray}
&&P_1=c_{\beta-\alpha}A+s_{\beta-\alpha}G_Z, \\
&&P_2=-s_{\beta-\alpha}A+c_{\beta-\alpha}G_Z.
\end{eqnarray}
In conventional linear gauges, only the mixing terms $W-G_W$ and
$Z-G_Z$ are removed from the theory. In contrast, our gauge-fixing
procedure also removes the unphysical vertices $WG_W\gamma$,
$WG_WZ$, $HWG_W\gamma$, $hWG_W\gamma$, $G_ZWG_W\gamma$, $HWG_WZ$,
$hWG_WZ$, and $G_ZWG_WZ$. In addition,  the unphysical vertices
$HWG_W$, $hWG_W$, and $G_ZWG_W$ are modified. The Feynman rules for the modified vertices are given in Table \ref{TABLE2}. Once again, note that
the couplings involving only physical scalars are not modified by
the gauge-fixing procedure.

\begin{table}
\caption{\label{TABLE2}Nonphysical coupling of the Higgs--Kinetic energy term that are affected by the nonlinear gauge--fixing procedure. Notices that the vertices which vanish in this gauge but not in conventional ones are indicated. In this Table, $\phi_a$ stands for $h$ o $H$.}
\begin{tabular}{|l|l|l|l|}
\hline Coupling & Vertex Function & Coupling & Vertex Function  \\
\hline $H(k_1)W_\mu^\pm(k_2)G_W^\mp(k_3)$&$\mp
igk_{1\mu}c_{\beta-\alpha}$&$h(k_1)W_\mu^\pm(k_2)G_W^\mp(k_3)$&$\mp
igk_{1\mu}s_{\beta-\alpha}$\\
\hline $G_Z(k_1)W_\mu^\pm(k_2)G_W^\mp(k_3)$&$ gk_{1\mu}$&$V_\nu W_\mu^\pm G_W^\mp$&$0$\\
\hline $V_\mu W_\nu^\pm\phi_a G_W^\mp$&0&$V_\mu W_\nu^\pm G_Z
G_W^\mp$&$0$\\
\hline
\end{tabular}
\end{table}

\subsection{The ghost sector}

Let us now discuss the structure of the ghost sector. It is
convenient to introduce the following definitions for the ghost
fields
\begin{eqnarray}
&&C^\pm =\frac{1}{\sqrt{2}}(C^1\mp i C^2),\\
&&C^Z=c_WC^3-s_WC, \\
&&C^A=s_WC^3+c_WC,
\end{eqnarray}
and similar expressions for the antighost fields. We can then write
the corresponding Lagrangian as follows:
\begin{equation}
{\cal L}_F={\cal L}_{FV1}+{\cal L}_{FS1}+{\cal L}_{FV2}+{\cal
L}_{FS2}+{\cal L}_{F3},
\end{equation}
where
\begin{eqnarray}
{\cal L}_{FV1}&=&-\bar{C}^-(\delta f^+_V)-\bar{C}^+(\delta f^-_V)
-\bar{C}^Z(\delta f^Z_V)-\bar{C}^A(\delta f^A_V),\label{g1}\\
{\cal L}_{FS1}&=&-\bar{C}^-(\delta f^+_S)-\bar{C}^+(\delta f^-_S)
-\bar{C}^Z(\delta f^Z_S),\label{g2}
\end{eqnarray}
\begin{eqnarray}
{\cal
L}_{FV2}&=&-\frac{2ig}{\xi}\Big[(f^-_V\bar{C}^+-f^+_V\bar{C}^-)(c_WC^Z+s_WC^A)\nonumber
\\
&&+(c_W\bar{C}^Z+s_W\bar{C}^A)(f^+_VC^--f^-_V
C^+)+(c_Wf^Z_V+s_Wf^A_V)(\bar{C}^-C^+-\bar{C}^+C^-)\Big],\label{g3}\\
{\cal
L}_{FS2}&=&-\frac{2ig}{\xi}\Big[(f^-_S\bar{C}^+-f^+_S\bar{C}^-)(c_WC^Z+s_WC^A)\nonumber
\\
&&+(c_W\bar{C}^Z+s_W\bar{C}^A)(f^+_SC^--f^-_S
C^+)+c_Wf^Z_S(\bar{C}^-C^+-\bar{C}^+C^-)\Big]\label{g4},
\end{eqnarray}
\begin{eqnarray} \label{LG4}
{\cal
L}_{F3}&=&2g^2\Big(1-\frac{2}{\xi}\Big)\Big[\bar{C}^+\bar{C}^-C^+C^-\nonumber \\
&&+
(c_W\bar{C}^Z+s_W\bar{C}^A)(\bar{C}^+C^-+\bar{C}^-C^+)(c_WC^Z+s_WC^A)\Big].
\end{eqnarray}
It is important to point out that the terms ${\cal L}_{FV2}$, ${\cal
L}_{FS2}$, and ${\cal L}_{F3}$ are not present when the FPM is used.
The variations of the charged gauge-fixing functions are given by
\begin{eqnarray}
&&\delta f^+_V=\bar{D}_\mu \hat{D}^\mu
C^++ig(c_WC^Z+s_WC^A)(\bar{D}_\mu
W^{+\mu})+\frac{ig}{c_W}W^{+\mu}(\partial_\mu C^Z), \\
&&\delta
f^+_S=\frac{g^2\xi}{2}\Big\{\Big[G^-_WG^+_W-\frac{1}{2}(\varphi^{02}+G^2_Z)\Big]
C^+-(\varphi^0-iG_Z)G^+_W(c_WC^Z+s_WC^A)\Big\},
\end{eqnarray}
where $\delta f^-_{V,S}=(\delta f^+_{V,S})^\dag$. These functions
transform covariantly under the $U_e(1)$ group. As for the
variations of the neutral functions, they are given by
\begin{eqnarray}
&&\delta f^Z_V=\Box C^Z+igc_W\partial_\mu
(W^{-\mu}C^+-W^{+\mu}C^-),\\
&&\delta
f^Z_S=\frac{g\xi}{2}m_Z(G^+_WC^-+G^-_WC^+)-\frac{g\xi}{2c_W}
m_Z\varphi^{0}C^Z, \\
&&\delta f^A_V=\Box C^A+ie\partial_\mu (W^{-\mu}C^+-W^{+\mu}C^-).
\end{eqnarray}
We can see that the complete ghost sector is invariant under the
$U_e(1)$ group. This gives rise to some vertices which are not
present in a linear gauge. For instance, the vertex $\bar{C}^\pm
C^\mp \gamma \gamma$ is a direct consequence of $U_e(1)$-gauge
invariance. It is clear thus that the ghost fields satisfy QED-like
Ward identities, which can simplify considerably some loop
calculations. Since this sector is strongly affected by the nonlinear gauge--fixing procedure, we present all the Feynman rules. They are given in Tables \ref{TABLE3}, \ref{TABLE4}, \ref{TABLE5}, and \ref{TABLE6}. In these Tables, all the momenta are token incoming.

\begin{table}
\caption{\label{TABLE3}Trilinear couplings involving ghost fields.}
\begin{tabular}{|l|l|}
\hline Coupling & Vertex Function   \\
\hline $\bar C^\pm C^\mp H$&$igm_W\xi c_{\beta-\alpha}$\\
$\bar{C}^\pm C^\mp h$&$igm_W\xi s_{\beta-\alpha}$\\
\hline $\bar C^\pm C^\mp G_Z$&$\pm 2g m_Zc_W$\\
$\bar C^Z C^\pm G_W^\mp$&$\frac{-igm_Z\xi}{2}$\\
\hline $\bar C^Z C^Z H$&$\frac{igm_Z\xi}{2c_W}c_{\beta-\alpha}$\\
$\bar C^Z C^Z h$&$\frac{igm_Z\xi}{2c_W}s_{\beta-\alpha}$\\
\hline $\bar C^\pm(\bar P)C^\mp (P)A_\mu(k)$&$\pm
igs_W\left[\left(1+\frac{2}{\xi}\right)P-\left(1-\frac{2}{\xi}\right)\bar
P\right]_\mu$\\
$\bar C^\pm(\bar P)C^\mp (P)Z_\mu(k)$&$\mp igc_W\left[\left(t_W^2-\frac{2}{\xi}\right)P+\left(1-\frac{2}{\xi}\right)\bar P\right]_\mu$\\
\hline $\bar C^A(\bar P)C^\pm(P)W_{\mu}^\mp (k)$&$\mp
igs_W\left[\left(1-\frac{2}{\xi}\right)\bar
P-\frac{2}{\xi}P\right]_\mu$\\
$\bar C^\pm(\bar P)C^A(P)W_\mu^\mp(k)$&$\pm igs_W\left(1-\frac{2}{\xi}\right)\left(\bar P+P\right)_\mu$\\
\hline $\bar C^Z(\bar P)C^\pm(P)W_{\mu}^\mp (k)$&$\mp
igc_W\left[\left(1-\frac{2}{\xi}\right)\bar
P-\frac{2}{\xi}P\right]_\mu$\\
$\bar C^\pm(\bar P)C^Z(P)W_\mu^\mp(k)$&$\mp i g
c_W\left[\left(t_W^2+\frac{2}{\xi}\right)P-\left(1-\frac{2}{\xi}\right)\bar
P\right]_\mu$\\
\hline
\end{tabular}
\end{table}

\begin{table}
\caption{\label{TABLE4}Couplings involving four ghosts. This class of couplings are arise as a consequence of using the BRST formalism.}
\begin{tabular}{|l|l|l|l|}
\hline Coupling & Vertex Function & Coupling & Vertex Function  \\
\hline $\bar C^+\bar C^- C^+ C^-$&$2ig^2(1-\frac{2}{\xi})$&&\\
\hline $\bar C^Z \bar C^\pm C^\mp
C^A$&$2ig^2(1-\frac{2}{\xi})s_Wc_W$&$\bar C^A \bar C^\pm C^\mp
C^Z$&$2ig^2(1-\frac{2}{\xi})s_Wc_W$\\
\hline $\bar C^A \bar C^\pm C^\mp
C^A$&$2ig^2(1-\frac{2}{\xi})s_W^2$&$\bar C^Z \bar C^\pm C^\mp
C^Z$&$2ig^2(1-\frac{2}{\xi})c_W^2$\\
\hline
\end{tabular}
\end{table}

\begin{table}
\caption{\label{TABLE5}Couplings generated by two ghosts and two scalar fields. This class of couplings are absent in conventional linear gauges.}
\begin{tabular}{|l|l|l|l|}
\hline Coupling & Vertex Function & Coupling & Vertex Function  \\
\hline $\bar C^\pm C^A G_W^\mp H$&$\frac{ig^2}{2}(\xi-2)s_W
c_{\beta-\alpha}$&$\bar C^A C^\pm G_W^\mp H$&$-ig^2s_W
c_{\beta-\alpha}$\\
\hline $\bar C^\pm C^A G_W^\mp h$&$\frac{ig^2}{2}(\xi-2)s_W
s_{\beta-\alpha}$&$\bar C^A C^\pm G_W^\mp h$&$-ig^2s_W
s_{\beta-\alpha}$\\
\hline $\bar C^\pm C^A G_W^\mp G_Z$&$\mp\frac{g^2}{2}(\xi-2)s_W
$&$\bar C^A C^\pm G_W^\mp G_Z$&$\pm g^2s_W $\\
\hline $\bar C^\pm C^Z G_W^\mp H$&$\frac{ig^2}{2}(\xi-2)c_W
c_{\beta-\alpha}$&$\bar C^Z C^\pm G_W^\mp H$&$-ig^2c_W
c_{\beta-\alpha}$\\
\hline $\bar C^\pm C^Z G_W^\mp h$&$\frac{ig^2}{2}(\xi-2)c_W
s_{\beta-\alpha}$&$\bar C^Z C^\pm G_W^\mp h$&$-ig^2c_W
s_{\beta-\alpha}$\\
\hline $\bar C^\pm C^Z G_W^\mp G_Z$&$\mp\frac{g^2}{2}(\xi-2)c_W
$&$\bar C^Z C^\pm G_W^\mp G_Z$&$\pm g^2c_W $\\
\hline $\bar C^\pm C^\mp
HH$&$\frac{ig^2\xi}{2}c_{\beta-\alpha}^2$&$\bar C^\pm C^\mp
hh$&$\frac{ig^2\xi}{2}s_{\beta-\alpha}^2$\\
\hline $\bar C^\pm C^\mp
Hh$&$\frac{ig^2\xi}{2}c_{\beta-\alpha}s_{\beta-\alpha}$&$\bar
C^\pm C^\mp G_Z G_Z$&$\frac{ig^2\xi}{2}$\\
\hline$\bar C^\pm C^\mp G_W^\pm G_W^\mp$&$-\frac{ig^2\xi}{2}$&&\\
\hline
\end{tabular}
\end{table}

\begin{table}
\caption{\label{TABLE6}Quartic couplings generated by two ghosts and two vector bosons. This class of couplings are absent in conventional linear gauges.}
\begin{tabular}{|l|l|l|l|}
\hline Coupling & Vertex Function & Coupling & Vertex Function  \\
\hline$\bar C^\pm C^\mp Z_\mu A_\nu$&$ig^2t_Wc_{2W}g_{\mu\nu}$&&\\
\hline $\bar C^\pm C^\mp A_\mu A_\nu$&$2ig^2s_W^2g_{\mu\nu}$
&$\bar C^\pm C^\mp Z_\mu Z_\nu$&$-2ig^2s_W^2g_{\mu\nu}$\\
\hline $\bar C^A C^\pm A_\mu W_\nu^\mp$&$\frac{-2i}\xi g^2
s_W^2g_{\mu\nu}$&$\bar C^\pm C^A
A_{\mu}W_\nu^\mp$&$\frac{-i(\xi-2)}{\xi}g^2s_W^2g_{\mu\nu}$\\
\hline $\bar C^A C^\pm Z_\mu W_\nu^\mp$&$\frac{2i}\xi
g^2s_W^2t_Wg_{\mu\nu}$&$\bar C^\pm C^A
Z_{\mu}W_\nu^\mp$&$\frac{i(\xi-2)}{\xi}g^2s_W^2t_Wg_{\mu\nu}$\\
\hline $\bar C^Z C^\pm Z_\mu W_\nu^\mp$&$\frac{2i}\xi
g^2s_W^2g_{\mu\nu}$&$\bar C^\pm C^Z
Z_{\mu}W_\nu^\mp$&$\frac{i(\xi-2)}{\xi}g^2s_W^2g_{\mu\nu}$\\
\hline $\bar C^Z C^\pm A_\mu W_\nu^\mp$&$\frac{-2i}\xi
g^2s_Wc_Wg_{\mu\nu}$&$\bar C^\pm C^Z
A_{\mu}W_\nu^\mp$&$\frac{-i(\xi-2)}{\xi}g^2s_Wc_Wg_{\mu\nu}$\\
\hline
\end{tabular}
\end{table}

\section{Summary}
\label{c}In this paper, we have presented a nonlinear $R_\xi$ gauge for the two Higgs doublet model based in the BRST symmetry and covariance under the electromagnetic gauge group. This gauge--fixing procedure allows one to remove of the theory the unphysical vertices $WG_W\gamma$,
$WG_WZ$, $HWG_W\gamma$, $hWG_W\gamma$, $G_ZWG_W\gamma$, $HWG_WZ$,
$hWG_WZ$, and $G_ZWG_WZ$. Due to the covariance of the gauge--fixing procedure under the electromagnetic gauge group, all the charged particles of the model satisfy simple Ward identities, which greatly simplifies the loop calculations. This type of gauges are particulary useful in calculating loop processes that involves at least an external photon. The main advantages are: (a) the number of Feynman diagrams for a given process reduces considerably with respect to those appearing in conventional linear gauges, (b) each type of charged particle circulating in the loop generates by itself electromagnetic gauge invariance, and (c) for those one--loop processes that are free of ultraviolet divergences, such as interactions among neutral particles and photons, the cancelation of divergences also occurs in a simple way through of subsets of diagrams involving the same type of charged particle.  The Feynman rules for the new vertices arising in this gauge, as well as for those already present in conventional linear gauge but that are modified by this gauge--fixing procedure, are presented.


\end{document}